# Probing the band splitting near the Γ point in the van der Waals magnetic semiconductor CrSBr

Kaiman Lin[1,2], Yi Li[2,3], Mahdi Ghorbani-Asl[2], Zdenek Sofer[4], Stephan Winnerl[2], Artur Erbe[2,3], Arkady V. Krasheninnikov[2], Manfred Helm[2,3], Shengqiang Zhou[2], Yaping Dan[1,*], Slawomir Prucnal[2,*]

[1] University of Michigan-Shanghai Jiao Tong University Joint Institute, Shanghai Jiao Tong University, 20024 Shanghai, P. R. China

[2] Helmholtz-Zentrum Dresden-Rossendorf, Institute of Ion Beam Physics and Materials Research, Bautzner Landstrasse 400, 01328 Dresden, Germany

[3] TU Dresden, 01062 Dresden, Germany

[4] Department of Inorganic Chemistry, University of Chemistry and Technology Prague, Technická 5, 16628 Prague 6, Czech Republic

[*]corresponding authors: s.prucnal@hzdr.de, yaping.dan@sjtu.edu.cn

**ABSTRACT:** This study investigates the electronic band structure of Chromium Sulfur Bromide (CrSBr) through comprehensive photoluminescence (PL) characterization. We clearly identify low-temperature optical transitions between two closely adjacent conduction-band states and two different valence-band states. The analysis of the PL data robustly unveils energy splittings, bandgaps and excitonic transitions across different thicknesses of CrSBr, ranging from monolayer to bulk. Temperature-dependent PL measurements elucidate the stability of the band splitting below the Néel temperature, suggesting that magnons coupled with excitons are responsible for the symmetry breaking and brightening of the transitions from the secondary conduction band minimum (CBM2) to the global valence band maximum (VBM1). Collectively, these results not only reveal band splitting in both the conduction and valence bands, but also point to an intricate interplay between the optical, electronic and magnetic properties of antiferromagnetic two-dimensional van der Waals crystals.

**KEYWORDS:** *CrSBr, antiferromagnetic semiconductor, van der Waals materials, band splitting*

Graphic abstract

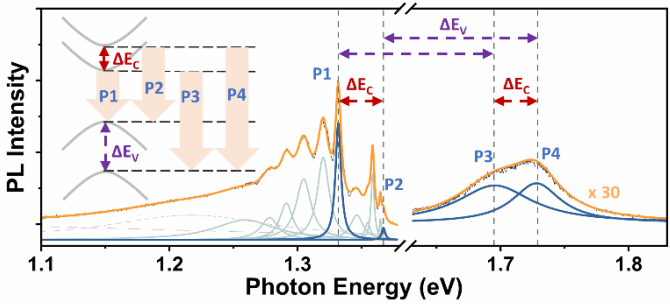

CrSBr, an A-type antiferromagnetic (AFM) material, features van der Waals ferromagnetic monolayers with antiferromagnetic coupling along the stacking direction.[1,2] Distinct from conventional magnetic materials, CrSBr is a semiconductor with a sizable direct bandgap, approximately 1.5 eV.[1,3] Additionally, it is characterized by a high transition temperature for antiferromagnetic coupling, reaching up to 132 K in its bulk form.[1,4,5] These exceptional properties make CrSBr a material of great interest for optoelectronics, spintronics, and quantum technology applications.[6–9]

Understanding the band structure of CrSBr is critical for comprehending its physical properties and potential in optoelectronic applications. By employing scanning tunneling spectroscopy (STS) and PL measurements at 300 K, Telford et al.[4] demonstrated that CrSBr is a direct bandgap semiconductor, with an electronic bandgap $E_g$ = 1.5 ± 0.2 eV and an excitonic PL peak at 1.25 ± 0.07 eV. Linhart et al.[10] utilized Hubbard-corrected density functional theory (DFT+U) calculations alongside calculations with hybrid functional HSE06, revealing an antiferromagnetic ground state for bulk CrSBr. In addition, these ab-initio calculations could yield different bandgap values of 0.85 eV and 1.85 eV depending on the calculation model. The photoreflectance and absorption measurements at 20 K disclosed three photoreflectance resonances at 1.39 eV, 1.45 eV and 1.77 eV, which can be associated with direct transitions in CrSBr.[10] Wilson et al.[11] carried out electronic structure calculations of monolayer CrSBr in its ferromagnetic (FM) ground state within GW approximation for the self energy, identifying a bandgap of approximately 1.8 eV with highly anisotropic band dispersion. The polarization-resolved differential reflectance (ΔR/R) spectra of bilayer CrSBr revealed the polarization-dependent excitonic transitions at 1.34 eV and 1.75 eV. The lower energy emission was identified as the direct bandgap transition, while the higher energy emission could be attributed to the transition from the global conduction band minimum (CBM1) to the secondary valence band maximum (VBM2) situated about 0.4 eV below the global valence band maximum (VBM1). Additionally, a PL peak splitting by roughly 40 meV was experimentally observed in the AFM phase for CrSBr flakes thicker than bilayer. The dipole forbidden optical transition between secondary conduction band minimum (CBM2) and VBM1 could be brightened under a reduction in symmetry, for instance, by an asymmetric dielectric environment. In addition, it was proposed that there is no significant thickness dependence in the calculated band structure. In contrast, Klein et al.[12] reported varying energy splitting at the Γ point for different CrSBr thicknesses, from 33 meV for a monolayer (1L), to 168 meV for bulk. However, these theoretical predictions of thickness-dependent band splitting in the conduction band of CrSBr have not been experimentally verified.

Our research bridges this gap through a series of experimental findings, providing new insights into band splitting in both valence band and conduction band of CrSBr. We have identified band splitting in the conduction band, which is manifested in both higher- and lower-energy emissions. Notably, our study reveals a consistent pattern of band splitting across a wide range of CrSBr thicknesses, from monolayer to bulk. This uniformity in band splitting at the Γ-point, regardless of CrSBr thickness, offers important insights into the intrinsic properties of this material and reinforces its potential for diverse technological applications.

## RESULTS AND DISCUSSION

The schematic representation of band splitting in both the conduction band and valence band near the Γ point in the band structure of CrSBr is shown in Figure 1a, where P1 represents the transition from CBM1 to VBM1, P2 for the transition from CBM2 to VBM1, P3 for CBM1 to VBM2, and P4 for CBM2 to VBM2. The following experimental analysis reveals that the splitting within the conduction band, $\Delta E_C$, is approximately 35 meV, while the splitting within the valence band, $\Delta E_V$, is around 0.4 eV. It is worth noting that the value for $\Delta E_C$ remains the same for various thicknesses of CrSBr.

Figure 1b displays the photoluminescence (PL) spectrum of a 103 nm thick CrSBr flake on a $SiO_2$/Si substrate, recorded at 300 K. The spectrum features two distinct emission peaks, positioned at around 1.30 eV and 1.73 eV. It is important to note that at 300 K, thermal broadening of the electronic states can merge closely spaced energy levels into a single peak. Consistent with the results of recent studies[11,12], the PL peak at approximately 1.30 eV is associated with the excitonic transition from the conduction band minimum to VBM1, which comprises P1 and P2 in Figure 1a. The higher-energy peak at 1.73 eV is interpreted as corresponding to the transition from the conduction band minimum to VBM2, which comprises P3 and P4 in Figure 1a. To confirm the ubiquity of this higher-energy peak, extensive PL measurements were performed on CrSBr flakes of varying thicknesses. As illustrated in Figure 1c, the higher-energy peak is consistently present in the PL spectra of CrSBr flakes with different thicknesses, confirming the band splitting in the valence band near the Γ point, $\Delta E_V$, at around 0.4 eV.

In a further investigation of the CrSBr band structure, PL measurement were conducted at 4 K across a range of thickness, from 1L to 103 nm. These results, depicted in Figure 1d, uniformly exhibit the most intense emission at approximately 1.332 eV, marked as peak P1. Notably, for thicknesses in the range from 8 nm to 103 nm, satellite peaks were observed on the lower energy side of peak P1. Based on our recent research,[13] these satellite peaks are attributed to exciton-phonon coupling states. Dirnberger *et al.*[14] have proposed the existence of exciton-photon coupling within CrSBr flakes across a broad thickness range, roughly from 10 to 1000 nm, through simulation. This coupling leads to exciton-polariton states, facilitated by the large oscillator strength of excitons in CrSBr, and exhibits a thickness-dependent behavior. Their studies have verified the existence of exciton-polariton states in CrSBr flakes, with the thinnest one being 75 nm. Here, we identify additional peaks, labeled 'E', in the PL spectra of CrSBr with thicknesses starting from 30 nm. These 'E' peaks likely arise from exciton-photon coupling, as evidenced by their variation across different flake thicknesses. Notably, in thinner flakes, the 'E' peak is absent, a phenomenon that could be attributed to the reduced oscillator strength of excitons and weaker photon confinement within thinner CrSBr flakes.[14] Furthermore, each spectrum includes a peak at around 1.367 eV, referred to as peak P2. The Lorentzian fitting of the PL spectra for the 30 nm, 45 nm, and 103 nm thick CrSBr, detailed in Figure S1a and Figure S2, further clarifies this observation. The energy difference between peaks P1 and P2, denoted by $\Delta E_{C1}$, quantified at around 35 meV, closely matches the reported theoretical energy gap[11] between CBM1 and CBM2. Based on these observations, we conclude that peak P1 corresponds to the transition from VBM1 to CBM1, and peak P2 to the transition from VBM1 to CBM2, with the splitting between VBM1 and VBM2, $\Delta E_C$, to be around 35 meV.

Notably, the positions for peak P1 remain constant for various thicknesses, from 1 L to 103 nm. This observation indicates that the optical bandgap of CrSBr maintains uniformity across these different thicknesses. In contrast, most two-dimensional (2D) materials, such as transition metal dichalcogenides (TMDCs), exhibit a significant change in bandgap when going from bulk to monolayer.[15–18] This phenomenon is primarily attributed to quantum confinement effects and the modulation of the electronic environment with thickness.[19,20] However, recent studies[12,21] suggest that bulk CrSBr can be conceptualized as a stack of weakly coupled monolayers that host quasi-1D excitons of high binding energy with strong quasiparticle interactions. The electronic properties of individual layers in CrSBr may be relatively unaffected by the presence or absence of adjacent layers due to weak interlayer interactions. Furthermore, antiparallel spin polarization in adjacent CrSBr layers may also contribute to this observed thickness-independent behavior of the bandgap.

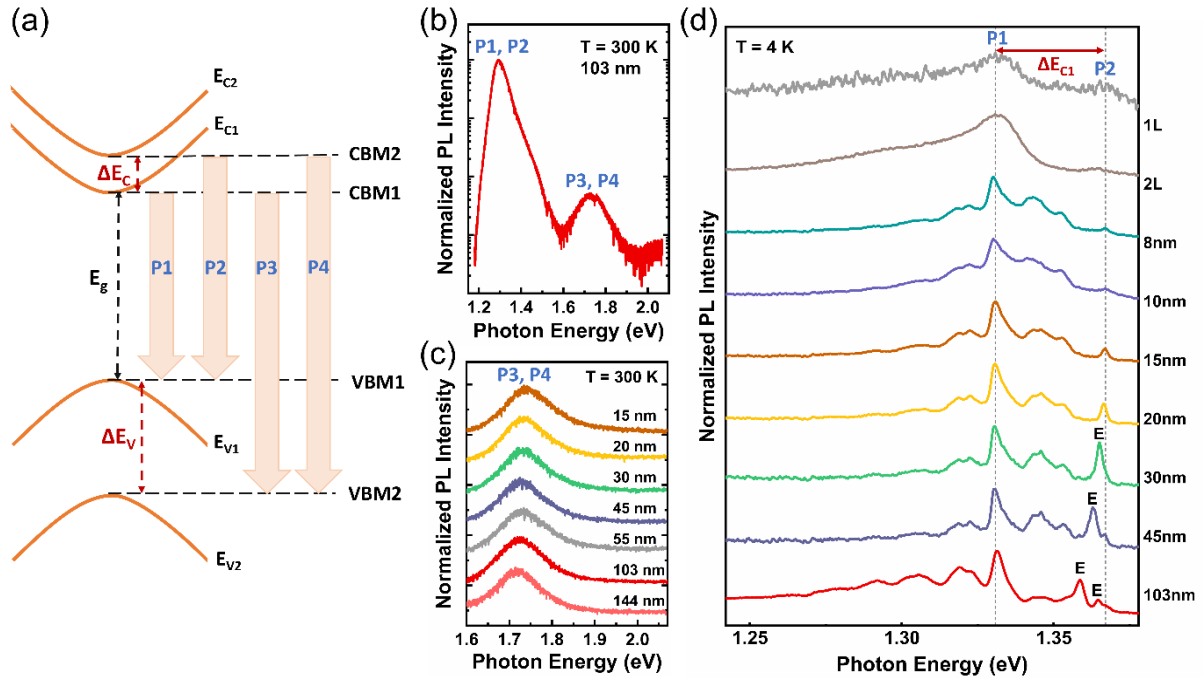

**Figure 1.** μ-PL spectra of thin CrSBr flakes. **(a)** Schematic representation of the band splitting in conduction band and valence band. **(b)** PL spectrum of 103 nm thick CrSBr flake on SiO$_2$/Si substrate acquired at 300 K, with excitation power of 10 mW. **(c)** PL spectra for CrSBr flakes with different thicknesses. **(d)** PL spectra for CrSBr flakes with different thicknesses at 4 K.

To explore the dynamics of conduction band splitting with respect to temperature, we carried out temperature-dependent PL measurement on CrSBr samples for different thicknesses: 1L, 2L (bilayer), and 103 nm. The evolution of the PL spectra with increasing temperature, from 4 K to 60 K, is illustrated in Figures 2a, 2b, and 2c for each respective thickness. Notably, peak P2 is observable within the range from 4 K to 30 K. However, as the temperature rises to 60 K, peak P2 becomes less distinguishable, likely due to thermal broadening effects.[22]

For 1L CrSBr, $\Delta E_{C1}$, the energy difference between peaks P1 and P2, consistently remains to be 34.5 meV when temperature increases from 4 K to 30 K. In the 2L CrSBr, this gap stands at 32.0 meV. The analysis of the 103 nm thick CrSBr, which is complicated by the presence of multiple peaks, necessitates a Lorentzian fitting approach to accurately determine the positions of peaks P1 and P2. The outcomes of this fitting for the 103 nm thick CrSBr at 4 K, 15 K and 30 K, are depicted in Figure S1. This fitting indicates that peaks P1 and P2 are separated by an energy gap of approximately 35.0±0.4 meV. The minor variation in the energy difference $\Delta E_{C1}$ across different thicknesses of CrSBr supports our hypothesis that the band splitting between CBM1 and CBM2, $\Delta E_C$, is around 35 meV and remains largely constant across different thicknesses. This observation of thickness-independent conduction-band splitting is corroborated by our computational findings, presented in Figure S3. Furthermore, it is important to highlight that this band splitting is stable at low temperatures and remains observable up to a temperature of 30 K.

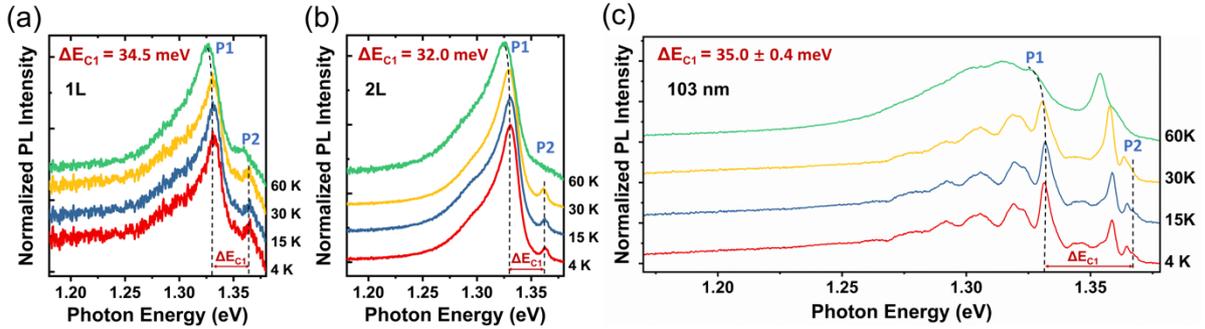

**Figure 2. Temperature-dependent µ-PL of CrSBr flake with different thicknesses.** Evolution of PL spectrum measured from T=4 K to T=60 K for **(a)** 1L, **(b)** 2L and **(c)** 103 nm thick CrSBr flake.

To further validate our hypothesis regarding the conduction band splitting, we traced the evolution of high energy peak in 103 nm thick CrSBr from 4 K to 300 K, as shown in Figure 3a. Intriguingly, a slight S-shaped shift in the peak position is observed with increasing temperatures, akin to behaviors reported in quantum wells and certain bulk III-V compound semiconductors.[23–27] In quantum wells, S-shaped temperature dependence of the PL peak energy is attributed to exciton localization in band-tail states, formed due to potential fluctuations resulting from variations in alloy composition at interfaces within quantum wells.[28] In the context of layered magnetic materials like CrSBr, potential fluctuations could arise from changes in magnetic coupling and the presence of strongly localized states caused by coupling of magnons with excitons. Telford *et al.*[4] highlighted the existence of FM states below 40 K and a transition from AFM to paramagnetic (PM) phase between 120 to 150 K, identifying a Néel temperature at around 132 K and observing short-range correlations that persist up to approximately 180 K.[14] Our research reveals deviations in the temperature-dependent peak behavior across specific temperature intervals: from 4 K to about 60 K, 60 K to around 175 K, and above 175 K to 300 K, aligning with different magnetic phase regimes in CrSBr.[10,13,14]

To enhance the precision in data analysis, Lorentzian fitting was applied to the spectra acquired across the temperature range under study. This fitting process for two representative PL spectra, collected at 4 K and 300 K, is detailed in Figure 3b and Figure 3c, respectively. Between 4 K to 60 K, two Lorentzian functions were necessary for efficient PL spectrum fitting. However, for temperatures starting from 100 K onwards, a single Lorentzian function proved to be sufficient for spectrum fitting, indicating the merging of closely spaced energy levels at elevated temperatures. The positions of the fitted peaks 'P3', 'P4' and composite 'P3, P4', along with their corresponding full width at half maximum (FWHM) versus temperature, are depicted in Figure 3d and Figure 3e, respectively. The energy gap between the fitted peaks P3 and P4, denoted by $\Delta E_{C2}$, is measured to be 35.0±2.7 meV between 4 K to 60 K. The congruence between $\Delta E_{C1}$ and $\Delta E_{C2}$ reinforces our hypothesis, affirming that the band splitting in the conduction band, $\Delta E_C$, is approximately 35 meV. Moreover, the increase in the FWHM of the peak 'P3, P4' with rising temperature corroborates the thermal broadening assumption.

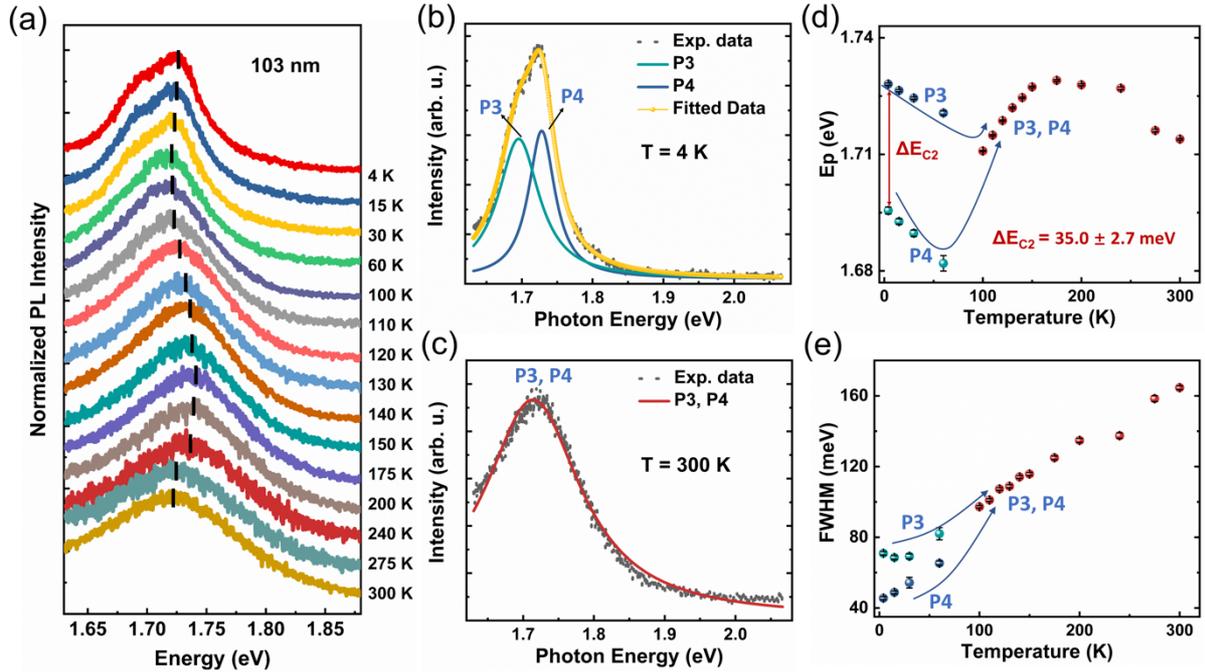

**Figure 3. Temperature-dependent μ-PL of 103 nm thick CrSBr flake. (a)** Evolution of PL spectrum measured from T = 4 K to T = 300 K. PL spectrum and Lorentzian Fitting curves of CrSBr acquired at **(b)** 4 K **(c)** 300 K, Dotted line: PL spectrum, Solid lines: individual peaks. Solid line with dots: sum of individual peaks. **(d)** Peak positions of peak 'P3', peak 'P4' and peak 'P3, P4' and **(e)** Corresponding FWHM obtained by Lorentzian Fitting of the PL spectra as a function of temperature.

To substantiate the excitonic nature of the high-energy emission in CrSBr, power-dependent PL measurements were carried out on the same sample at 4 K, as shown in Figure 4a. Notably, the normalized PL spectrum remained constant despite variations in pumping power, as depicted in Figure 4b. To delve deeper into the interactions between peaks P3 and P4, as well as the excitonic recombination process in CrSBr, we employed the Lorentzian fitting approach used for Figure 3b on the power-dependent PL spectra. Figure 4c showcases double-logarithmic plots illustrating the relationship between the PL intensity (I) of peaks P3 and P4 and the excitation power (L). This relationship adheres to a simple power law, expressed as

$$I = L^k, \qquad (1)$$

where $L$ is the excitation power and $k$ is a dimensionless exponential coefficient. Importantly, within the explored power range, we observed no saturation effects, and the PL intensity displayed an almost linear correlation with the excitation power. The exponential coefficient $k$ is instrumental in determining the deexcitation mechanisms in the material. A value of $k$ exceeding 1 suggests stimulated emission, whereas a value below 1 indicates the presence of non-radiative pathways, such as defects, Auger recombination, or exciton-to-trion conversion, phenomena often observed in 2D van der Waals crystals.[29–32] For the peaks P3 and P4 under investigation, the fitted exponent $k$ is approximately 1±0.1. This result is consistent with the radiative recombination of neutral excitons, corroborating our hypothesis that peak P3 is associated with the excitonic transition from CBM1 to VBM2, and peak P4 with the transition from CBM2 to VBM2.

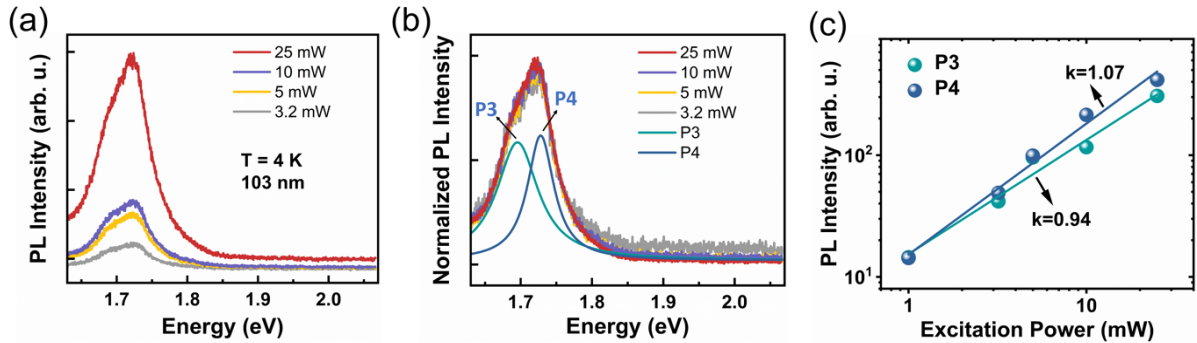

**Figure 4. Power-dependent μ-PL of 103 nm thick CrSBr flake. (a)** Power-dependent PL spectra at 4 K. **(b)** Normalized power-dependent PL spectra at 4 K. **(c)** PL intensity of peaks P3 and P4 versus excitation power at 4 K. Peak P3 and peak P4 were obtained by Lorentzian Fitting, as shown in Figure 4b.

The literature presents divergent findings, particularly from theoretical calculations, regarding the conduction band splitting in CrSBr. Experimental outcomes also vary widely due to different crystal growth techniques impacting crystal quality. Ideally, single-crystalline CrSBr exhibits antiferromagnetic properties with a Néel temperature of around 132 K. Recent studies indicate that native defects or disorder introduced by ion irradiation can trigger ferromagnetic states at temperatures typically below 100K, although the origin of ferromagnetic states remains unclear.[33] It is assumed that the presence of Br in interstitial or interlayer positions might locally enhance magnetic coupling between Cr atoms, altering the magnetic order from antiferromagnetic to ferromagnetic.[33] The optical characteristics of crystals are particularly sensitive to point defects. Our magnetometry analysis[33] suggests that our pristine CrSBr lacks low-temperature ferromagnetic phases, implying a defect-free material, thereby enabling a more detailed exploration of the electronic band structure through experimental optical methodologies. According to the selection rule, the transition from CBM2 to VBM1 is dipole forbidden.[11] However, certain environmental conditions can disrupt selection rules. The observation of radiative transitions from CBM2 exclusively below the Néel temperature implies that incoherent magnons, i.e., thermal fluctuations of the magnetic order, might play a major role.[13] We conjecture that magnons coupled with excitons are responsible for the exciton localization (manifested as an S-shaped peak position as a function of energy) and the reduction of symmetry that promotes the radiative transitions between CBM2 and VBM1.

**CONCLUSIONS**

To conclude, our investigation into the electronic band structure of CrSBr has uncovered band splitting in both the valence and conduction bands, revealed by low-temperature PL measurements. The conduction band splitting, evidenced by low and high energy emissions, is attributed to the transitions from CBM2 to VBM1 and from CBM2 to VBM2. Notably, thickness-dependent PL measurements show that this band splitting is uniform across a range of material thicknesses, from monolayer to bulk configurations. This universal behavior, independent of the CrSBr thickness, is crucial for our comprehension of its electronic structure. Furthermore, temperature-dependent PL studies demonstrate the stability of this band splitting below the Néel temperature, suggesting a pivotal role for magnon-exciton interactions in facilitating these symmetry-breaking phenomena and the resultant brightening of forbidden transitions. These findings substantially deepen our understanding of CrSBr, establishing a foundation for its future applications in the realms of optics, electronics, and magnetism, and expanding its potential technological impacts.

**METHODS**

**Crystal Growth and Sample Fabrication.** The synthesis of CrSBr bulk single crystals was conducted using a chemical vapor transport method. For this, high-purity chromium (99.99%, -60 mesh), sulfur (99.9999%, 1-6mm), and bromine (99.9999%) were mixed in a stoichiometric ratio of 1:1:1. The mixture was then sealed in a quartz ampoule under a high vacuum to maintain the integrity of the components. To prevent bromine evaporation, the ampoule was stored in a liquid nitrogen batch. Initially, a pre-reaction phase was conducted within the ampoule at 700°C for 20 hours, while keeping the other end at a lower temperature of 200°C. Afterward, the reacted mixture was transferred to a two-zone horizontal furnace for further processing. The furnace was initially set at 850°C at the source end and 900°C at the growth end. After 25 hours, the temperature gradient in the furnace was reversed, gradually increasing the hotter end from 880°C to 930°C over a period of 10 days. This procedure yielded high-quality CrSBr single crystals, which were carefully removed in an Ar glovebox. For flake samples, mechanical exfoliation using the scotch tape method was employed. The substrates used were silicon with a thermal oxide layer of 260 nm. The thickness of the exfoliated CrSBr flakes was determined using atomic force microscopy.

**Raman Spectroscopy.** Micro-Raman spectra were obtained using a 532 nm wavelength laser for excitation. The laser beam, focused on the sample through a 50× objective lens, produced a spot approximately 5 μm in diameter. The excitation power was consistently set at 1 mW. The scattered light, collected by the same objective, was subsequently dispersed using a Horiba LabRAM HR Evolution Raman spectrometer. Detection of the signal was carried out by a Charge-Coupled Device (CCD) camera, which was cooled with liquid nitrogen.

**Photoluminescence Spectroscopy.** Micro-PL spectra were obtained using a 532 nm wavelength laser for excitation. The laser beam, focused on the sample through a 50× objective lens, produced a spot approximately 5 μm in diameter. A closed-cycle helium cryostat was integrated into the micro-PL system to perform temperature-dependent measurements. These measurements were conducted under a base pressure below $1 \times 10^{-5}$ mbar. The PL signal was detected using either a CCD camera or an InGaAs camera, both cooled with liquid nitrogen.

## AUTHOR INFORMATION

**Author Contributions**

K.L. prepared the samples, conducted the AFM, Raman, and PL measurements, performed the data analysis and wrote the manuscript; Y.L. assisted with PL measurements; M.G. and A.K. performed the theoretical calculations; Z.S. synthesized CrSBr crystals; S.W. supported the Raman and PL measurements; S.Z., M.H. and A.E. discussed the theoretical issues and commented on the manuscript; S.P. and Y.D. supervised and directed the research.

**Notes**

The authors declare that they have no competing interests.

## ACKNOWLEDGMENTS

K. Lin gratefully acknowledges the financial support by Chinese Scholarship Council (File No. 202106230241). Y. Dan acknowledges the financial support by the special-key project of Innovation Program of Shanghai Municipal Education Commission (No. 2019-07-00-02-E00075), National Science Foundation of China (NSFC, No. 92065103), Oceanic Interdisciplinary Program of Shanghai Jiao Tong University (SL2022ZD107), Shanghai Jiao Tong University Scientific and Technological Innovation Funds (2020QY05). Z. Sofer was supported by ERC-CZ program (project LL2101) from Ministry of Education Youth and Sports (MEYS) and used large infrastructure from project reg. No. CZ.02.1.01/0.0/0.0/15_003/0000444 financed by the EFRR.

## ASSOCIATED CONTENT

**Supporting Information**

PL spectrum and Lorentzian Fitting curves of 103 nm thick CrSBr acquired at 4 K, 15 K and 30 K, PL spectrum and Lorentzian Fitting curves of 30 nm and 45 nm thick CrSBr acquired at 4 K. Atomic structure and electronic structure of CrSBr. Computational Details.

**REFERENCES**

(1) Göser, O.; Paul, W.; Kahle, H. G. Magnetic Properties of CrSBr. *J. Magn. Magn. Mater.* **1990**, *92* (1), 129–136. https://doi.org/10.1016/0304-8853(90)90689-N.

(2) Jiang, Z.; Wang, P.; Xing, J.; Jiang, X.; Zhao, J. Screening and Design of Novel 2D Ferromagnetic Materials with High Curie Temperature above Room Temperature. *ACS Appl. Mater. Interfaces* **2018**, *10* (45), 39032–39039. https://doi.org/10.1021/acsami.8b14037.

(3) Wang, C.; Zhou, X.; Zhou, L.; Tong, N.-H.; Lu, Z.-Y.; Ji, W. A Family of High-Temperature Ferromagnetic Monolayers with Locked Spin-Dichroism-Mobility Anisotropy: MnNX and CrCX (X = Cl, Br, I; C = S, Se, Te). *Sci. Bull.* **2019**, *64* (5), 293–300. https://doi.org/10.1016/j.scib.2019.02.011.

(4) Telford, E. J.; Dismukes, A. H.; Lee, K.; Cheng, M.; Wieteska, A.; Bartholomew, A. K.; Chen, Y.-S.; Xu, X.; Pasupathy, A. N.; Zhu, X.; Dean, C. R.; Roy, X. Layered Antiferromagnetism Induces Large Negative Magnetoresistance in the van Der Waals Semiconductor CrSBr. *Adv. Mater.* **2020**, *32* (37), 2003240. https://doi.org/10.1002/adma.202003240.

(5) Guo, Y.; Zhang, Y.; Yuan, S.; Wang, B.; Wang, J. Chromium Sulfide Halide Monolayers: Intrinsic Ferromagnetic Semiconductors with Large Spin Polarization and High Carrier Mobility. *Nanoscale* **2018**, *10* (37), 18036–18042. https://doi.org/10.1039/C8NR06368K.

(6) Burch, K. S.; Mandrus, D.; Park, J.-G. Magnetism in Two-Dimensional van Der Waals Materials. *Nature* **2018**, *563* (7729), 47–52. https://doi.org/10.1038/s41586-018-0631-z.

(7) Gong, C.; Zhang, X. Two-Dimensional Magnetic Crystals and Emergent Heterostructure Devices. *Science* **2019**, *363* (6428), eaav4450. https://doi.org/10.1126/science.aav4450.

(8) Li, H.; Ruan, S.; Zeng, Y.-J. Intrinsic Van Der Waals Magnetic Materials from Bulk to the 2D Limit: New Frontiers of Spintronics. *Adv. Mater.* **2019**, *31* (27), 1900065. https://doi.org/10.1002/adma.201900065.

(9) Mak, K. F.; Shan, J.; Ralph, D. C. Probing and Controlling Magnetic States in 2D Layered Magnetic Materials. *Nat. Rev. Phys.* **2019**, *1* (11), 646–661. https://doi.org/10.1038/s42254-019-0110-y.

(10) Linhart, W. M.; Rybak, M.; Birowska, M.; Scharoch, P.; Mosina, K.; Mazanek, V.; Kaczorowski, D.; Sofer, Z.; Kudrawiec, R. Optical Markers of Magnetic Phase Transition in CrSBr. *J. Mater. Chem. C* **2023**, *11* (25), 8423–8430. https://doi.org/10.1039/D3TC01216F.

(11) Wilson, N. P.; Lee, K.; Cenker, J.; Xie, K.; Dismukes, A. H.; Telford, E. J.; Fonseca, J.; Sivakumar, S.; Dean, C.; Cao, T.; Roy, X.; Xu, X.; Zhu, X. Interlayer Electronic Coupling on Demand in a 2D Magnetic Semiconductor. *Nat. Mater.* **2021**, *20* (12), 1657–1662. https://doi.org/10.1038/s41563-021-01070-8.

(12) Klein, J.; Pingault, B.; Florian, M.; Heißenbüttel, M.-C.; Steinhoff, A.; Song, Z.; Torres, K.; Dirnberger, F.; Curtis, J. B.; Weile, M.; Penn, A.; Deilmann, T.; Dana, R.; Bushati, R.; Quan, J.; Luxa, J.; Sofer, Z.; Alù, A.; Menon, V. M.; Wurstbauer, U.; Rohlfing, M.; Narang, P.; Lončar, M.; Ross, F. M. The Bulk van Der Waals Layered Magnet CrSBr Is a Quasi-1D Material. *ACS Nano* **2023**, *17* (6), 5316–5328. https://doi.org/10.1021/acsnano.2c07316.

(13) Lin, K.; Sun, X.; Dirnberger, F.; Li, Y.; Qu, J.; Wen, P.; Sofer, Z.; Söll, A.; Winnerl, S.; Helm, M.; Zhou, S.; Dan, Y.; Prucnal, S. Strong Exciton–Phonon Coupling as a Fingerprint of Magnetic Ordering in van Der Waals Layered CrSBr. *ACS Nano* **2024**. https://doi.org/10.1021/acsnano.3c07236.

(14) Dirnberger, F.; Quan, J.; Bushati, R.; Diederich, G. M.; Florian, M.; Klein, J.; Mosina, K.; Sofer, Z.; Xu, X.; Kamra, A.; García-Vidal, F. J.; Alù, A.; Menon, V. M. Magneto-Optics in a van Der Waals Magnet Tuned by Self-Hybridized Polaritons. *Nature* **2023**, *620* (7974), 533–537. https://doi.org/10.1038/s41586-023-06275-2.

(15) Mak, K. F.; Lee, C.; Hone, J.; Shan, J.; Heinz, T. F. Atomically Thin MoS$_2$: A New Direct-Gap Semiconductor. *Phys. Rev. Lett.* **2010**, *105* (13), 136805. https://doi.org/10.1103/PhysRevLett.105.136805.

(16) Kim, H.-C.; Kim, H.; Lee, J.-U.; Lee, H.-B.; Choi, D.-H.; Lee, J.-H.; Lee, W. H.; Jhang, S. H.; Park, B. H.; Cheong, H.; Lee, S.-W.; Chung, H.-J. Engineering Optical and Electronic Properties of WS$_2$ by Varying the Number of Layers. *ACS Nano* **2015**, *9* (7), 6854–6860. https://doi.org/10.1021/acsnano.5b01727.

(17) Jin, W.; Yeh, P.-C.; Zaki, N.; Zhang, D.; Sadowski, J. T.; Al-Mahboob, A.; van der Zande, A. M.; Chenet, D. A.; Dadap, J. I.; Herman, I. P.; Sutter, P.; Hone, J.; Osgood, R. M. Direct Measurement of the Thickness-Dependent Electronic Band Structure of MoS$_2$ Using Angle-Resolved Photoemission Spectroscopy. *Phys. Rev. Lett.* **2013**, *111* (10), 106801. https://doi.org/10.1103/PhysRevLett.111.106801.

(18) Splendiani, A.; Sun, L.; Zhang, Y.; Li, T.; Kim, J.; Chim, C.-Y.; Galli, G.; Wang, F. Emerging Photoluminescence in Monolayer MoS$_2$. *Nano Lett.* **2010**, *10* (4), 1271–1275. https://doi.org/10.1021/nl903868w.

(19) Kuc, A.; Zibouche, N.; Heine, T. Influence of Quantum Confinement on the Electronic Structure of the Transition Metal Sulfide TS$_2$. *Phys. Rev. B* **2011**, *83*, 245213. https://doi.org/10.1103/PhysRevB.83.245213.

(20) Meckbach, L.; Stroucken, T.; Koch, S. W. Influence of the Effective Layer Thickness on the Ground-State and Excitonic Properties of Transition-Metal Dichalcogenide Systems. *Phys. Rev. B* **2018**, *97* (3), 035425. https://doi.org/10.1103/PhysRevB.97.035425.

(21) Wu, F.; Gutiérrez-Lezama, I.; López-Paz, S. A.; Gibertini, M.; Watanabe, K.; Taniguchi, T.; von Rohr, F. O.; Ubrig, N.; Morpurgo, A. F. Quasi-1D Electronic Transport in a 2D Magnetic Semiconductor. *Adv. Mater.* **2022**, *34* (16), 2109759. https://doi.org/10.1002/adma.202109759.

(22) Qiao, H.; Abel, K. A.; van Veggel, F. C. J. M.; Young, J. F. Exciton Thermalization and State Broadening Contributions to the Photoluminescence of Colloidal PbSe Quantum Dot Films from 295 to 4.5 K. *Phys. Rev. B* **2010**, *82* (16), 165435. https://doi.org/10.1103/PhysRevB.82.165435.

(23) Minsky, M. S.; Fleischer, S. B.; Abare, A. C.; Bowers, J. E.; Hu, E. L.; Keller, S.; Denbaars, S. P. Characterization of High-Quality InGaN/GaN Multiquantum Wells with Time-Resolved Photoluminescence. *Appl. Phys. Lett.* **1998**, *72* (9), 1066–1068. https://doi.org/10.1063/1.120966.

(24) Chichibu, S.; Azuhata, T.; Sota, T.; Nakamura, S. Spontaneous Emission of Localized Excitons in InGaN Single and Multiquantum Well Structures. *Appl. Phys. Lett.* **1996**, *69* (27), 4188–4190. https://doi.org/10.1063/1.116981.

(25) Chichibu, S. F.; Azuhata, T.; Sugiyama, M.; Kitamura, T.; Ishida, Y.; Okumura, H.; Nakanishi, H.; Sota, T.; Mukai, T. Optical and Structural Studies in InGaN Quantum Well Structure Laser Diodes. *J. Vac. Sci. Technol. B Microelectron. Nanometer Struct. Process. Meas. Phenom.* **2001**, *19* (6), 2177–2183. https://doi.org/10.1116/1.1418404.

(26) Singh, S. D.; Porwal, S.; Sharma, T. K.; Rustagi, K. C. Temperature Dependence of the Lowest Excitonic Transition for an InAs Ultrathin Quantum Well. *J. Appl. Phys.* **2006**, *99* (6), 063517. https://doi.org/10.1063/1.2184431.

(27) Cho, Y.-H.; Gainer, G. H.; Fischer, A. J.; Song, J. J.; Keller, S.; Mishra, U. K.; DenBaars, S. P. "S-Shaped" Temperature-Dependent Emission Shift and Carrier Dynamics in InGaN/GaN Multiple Quantum Wells. *Appl. Phys. Lett.* **1998**, *73* (10), 1370–1372. https://doi.org/10.1063/1.122164.

(28) Dixit, V. K.; Porwal, S.; Singh, S. D.; Sharma, T. K.; Ghosh, S.; Oak, S. M. A Versatile Phenomenological Model for the S-Shaped Temperature Dependence of Photoluminescence Energy for an Accurate Determination of the Exciton Localization Energy in Bulk and Quantum Well Structures. *J. Phys. Appl. Phys.* **2014**, *47* (6), 065103. https://doi.org/10.1088/0022-3727/47/6/065103.

(29) Jung, J.-W.; Choi, H.-S.; Lee, Y.-J.; Taniguchi, T.; Watanabe, K.; Choi, M.-Y.; Jang, J. H.; Chung, H.-S.; Kim, D.; Kim, Y.; Cho, C.-H. Hexagonal Boron Nitride Encapsulation Passivates Defects in 2D Materials. arXiv October 3, 2022. https://doi.org/10.48550/arXiv.2210.00922.

(30) Schmidt, T.; Lischka, K.; Zulehner, W. Excitation-Power Dependence of the near-Band-Edge Photoluminescence of Semiconductors. *Phys. Rev. B* **1992**, *45* (16), 8989–8994. https://doi.org/10.1103/PhysRevB.45.8989.

(31) Kuechle, T.; Klimmer, S.; Lapteva, M.; Hamzayev, T.; George, A.; Turchanin, A.; Fritz, T.; Ronning, C.; Gruenewald, M.; Soavi, G. Tuning Exciton Recombination Rates in Doped Transition Metal Dichalcogenides. *Opt. Mater. X* **2021**, *12*, 100097. https://doi.org/10.1016/j.omx.2021.100097.

(32) Tongay, S.; Zhou, J.; Ataca, C.; Liu, J.; Kang, J. S.; Matthews, T. S.; You, L.; Li, J.; Grossman, J. C.; Wu, J. Broad-Range Modulation of Light Emission in Two-Dimensional Semiconductors by Molecular Physisorption Gating. *Nano Lett.* **2013**, *13* (6), 2831–2836. https://doi.org/10.1021/nl4011172.

(33) Long, F.; Ghorbani-Asl, M.; Mosina, K.; Li, Y.; Lin, K.; Ganss, F.; Hübner, R.; Sofer, Z.; Dirnberger, F.; Kamra, A.; Krasheninnikov, A. V.; Prucnal, S.; Helm, M.; Zhou, S. Ferromagnetic Interlayer Coupling in CrSBr Crystals Irradiated by Ions. *Nano Lett.* **2023**, *23* (18), 8468–8473. https://doi.org/10.1021/acs.nanolett.3c01920.

# Supporting Information

# Probing the band splitting near the Γ point in the van der Waals magnetic semiconductor CrSBr

Kaiman Lin[1,2], Yi Li[2,3], Mahdi Ghorbani-Asl[2], Zdenek Sofer[4], Stephan Winnerl[2], Artur Erbe[2,3], Arkady V. Krasheninnikov[2], Manfred Helm[2,3], Shengqiang Zhou[2], Yaping Dan[1,*], Slawomir Prucnal[2,*]

[1] University of Michigan-Shanghai Jiao Tong University Joint Institute, Shanghai Jiao Tong University, 20024 Shanghai, P. R. China

[2] Helmholtz-Zentrum Dresden-Rossendorf, Institute of Ion Beam Physics and Materials Research, Bautzner Landstrasse 400, 01328 Dresden, Germany

[3] TU Dresden, 01062 Dresden, Germany

[4] Department of Inorganic Chemistry, University of Chemistry and Technology Prague, Technická 5, 16628 Prague 6, Czech Republic

[*]corresponding authors: s.prucnal@hzdr.de, yaping.dan@sjtu.edu.cn

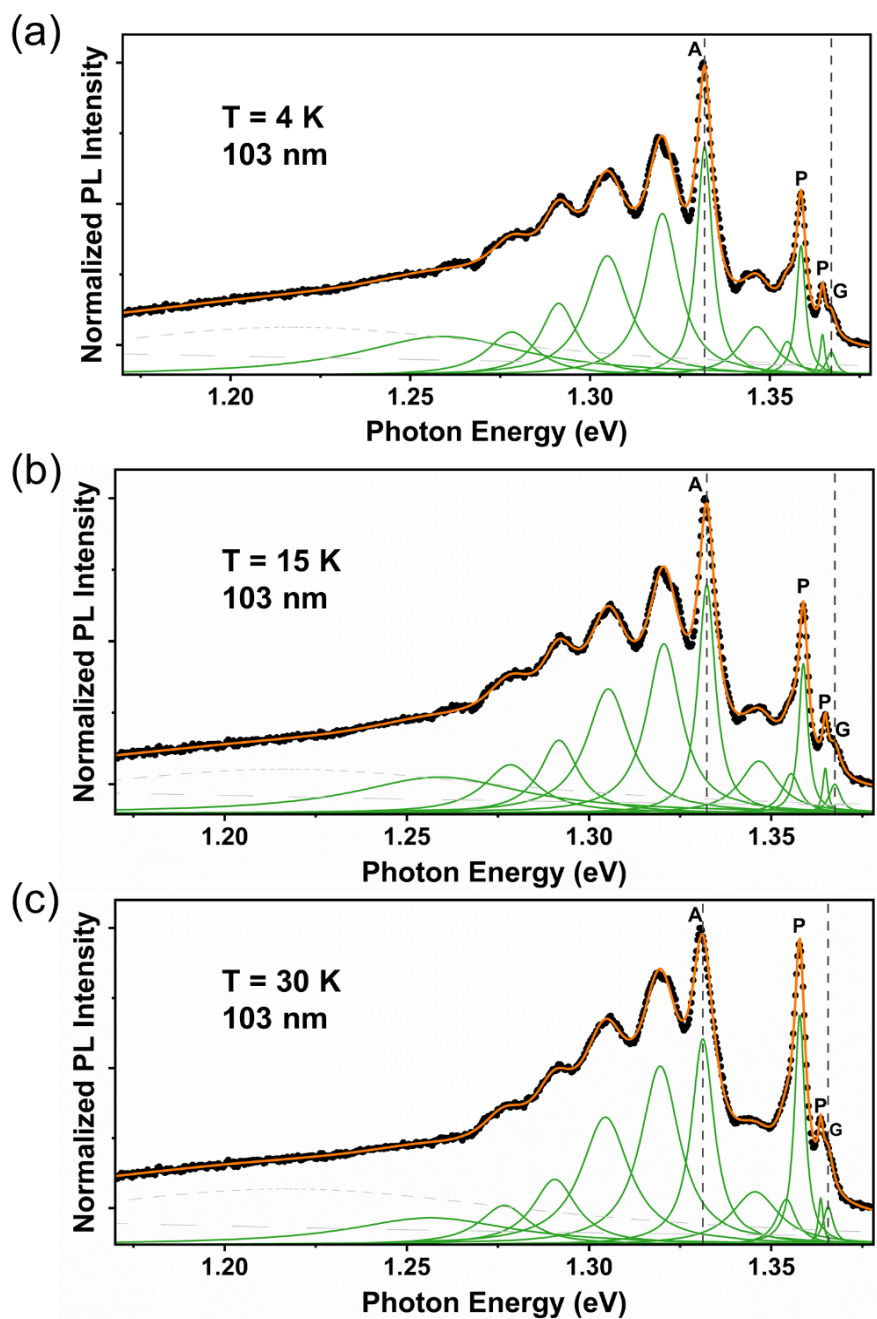

**Figure S1.** PL spectrum and Lorentzian Fitting curves of 103 nm thick CrSBr acquired at (a) 4 K, (b)15 K and (c) 30 K. Black dot: PL spectrum, Orange line: sum of individual peaks, Green lines: individual peaks.

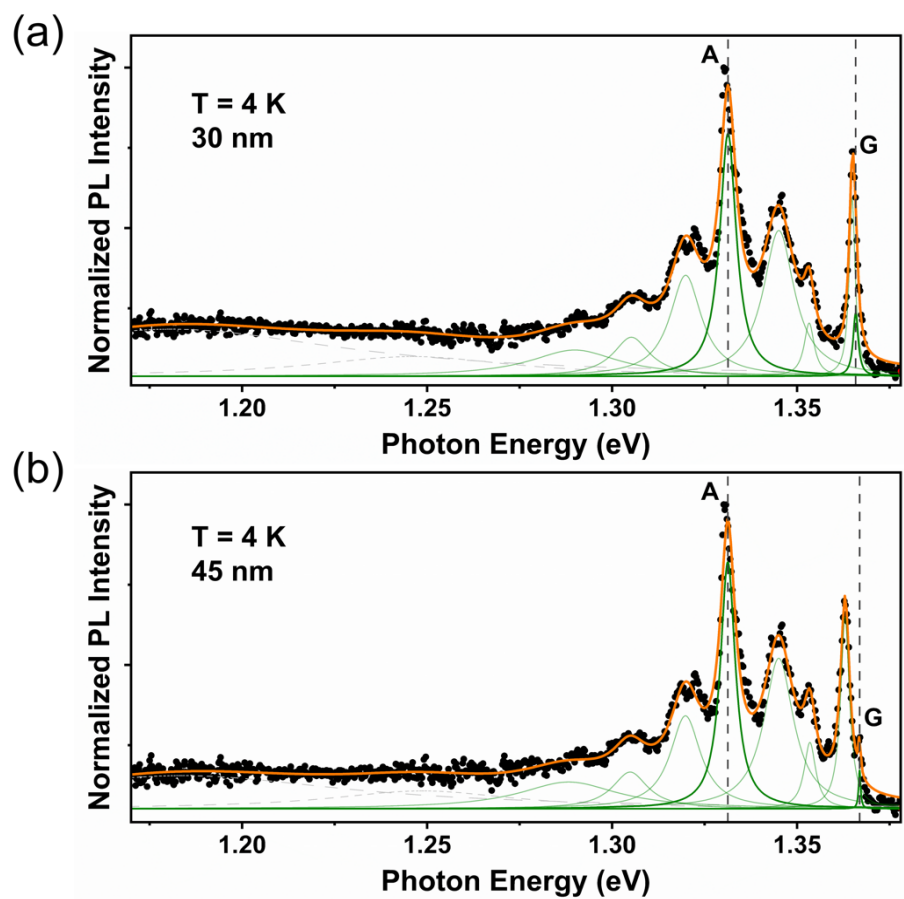

**Figure S2.** PL spectrum and Lorentzian Fitting curves of (a) 30 nm, (b) 45 nm thick CrSBr acquired at 4 K.

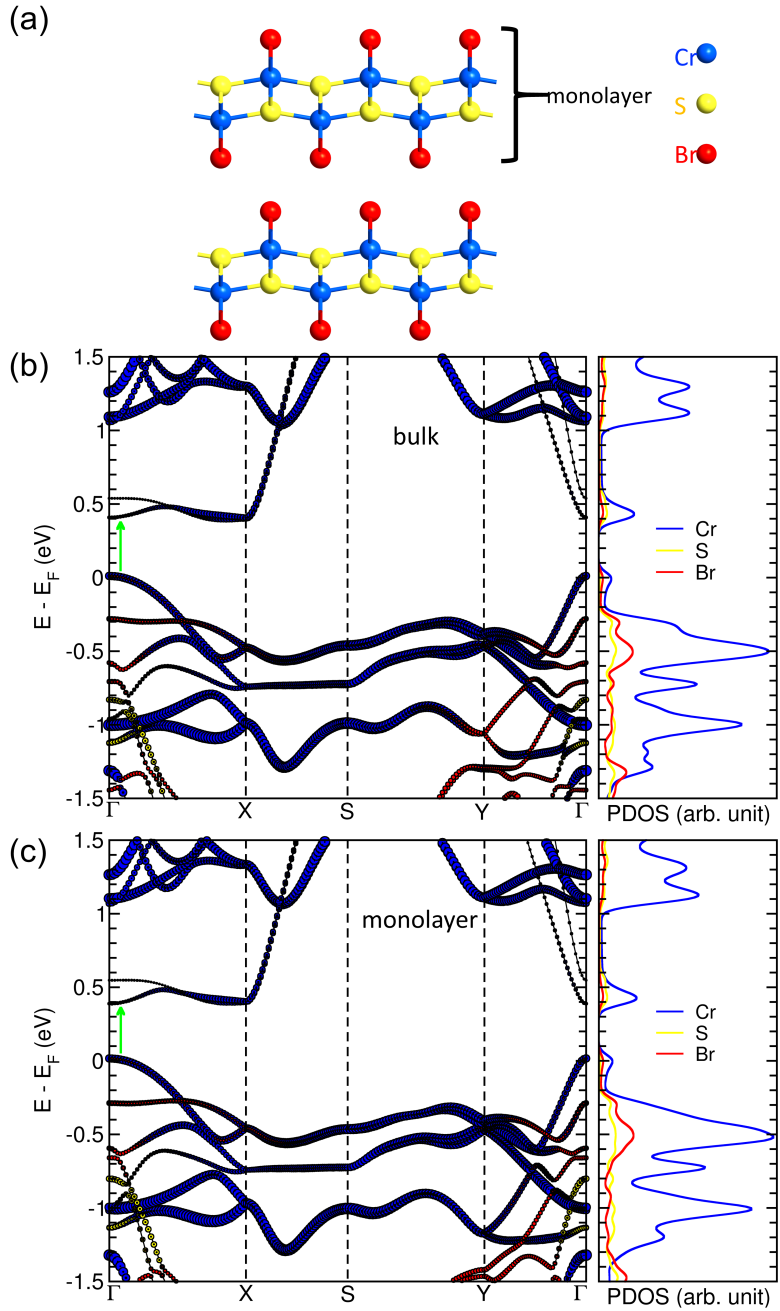

**Figure S3**. Atomic structure and electronic structure of CrSBr. (a) Atomic structure of CrSBr. Band structure and partial densities of states (PDOS) of (b) bulk and (c) monolayer CrSBr. PDOS indicates projections on the Cr, S and Br orbitals (blue, yellow, red). The calculations include spin-orbit interactions.

**Computational Details:** Spin-polarized density functional theory (DFT) calculations were carried out using Vienna Ab initio Simulation Package (VASP), based on the plane-wave projector augmented-wave (PAW) method.[1,2] The exchange-correlation functional was employed in the generalized gradient approximation of Perdew-Burke-Ernzerhof.[3] van der Waals (vdW) interactions were taken into account within the many-body dispersions method[4]. An energy cut-off of 600 eV was set for plane-wave expansion of all calculations chosen. The Brillouin zone of the system was sampled using 8 × 8 × 1 k-mesh for monolayer and 8× 8 × 8 k-mesh for bulk materials. The effect of spin-orbit coupling is included in the electronic structure calculations.

**Supporting references**

(1) Kresse, G.; Furthmüller, J. Efficiency of Ab-Initio Total Energy Calculations for Metals and Semiconductors Using a Plane-Wave Basis Set. *Comput. Mater. Sci.* **1996**, *6* (1), 15–50. https://doi.org/10.1016/0927-0256(96)00008-0.

(2) Kresse, G.; Furthmüller, J. Efficient Iterative Schemes for Ab Initio Total-Energy Calculations Using a Plane-Wave Basis Set. *Phys. Rev. B* **1996**, *54*, 11169. https://doi.org/10.1103/PhysRevB.54.11169.

(3) Perdew, J. P.; Burke, K.; Ernzerhof, M. Generalized Gradient Approximation Made Simple. *Phys. Rev. Lett.* **1996**, *77* (18), 3865–3868. https://doi.org/10.1103/PhysRevLett.77.3865.

(4) Tawfik, S. A.; Gould, T.; Stampfl, C.; Ford, M. J. Evaluation of van Der Waals Density Functionals for Layered Materials. *Phys. Rev. Mater.* **2018**, *2* (3), 034005. https://doi.org/10.1103/PhysRevMaterials.2.034005.